\documentstyle{elsart}

\begin{document}
\begin{frontmatter}
\title{Modelling of Infrared emission from Cyg X-3 and the UKIRT IRCAM3 point spread function}
\author{R.N. Ogley}
\address{Dept of Physics, The Open University, Milton Keynes, MK7 6AA, UK}
\author{S.J. Bell Burnell}
\address{Dept of Physics, The Open University, Milton Keynes, MK7 6AA, UK}
\author{R.P. Fender}
\address{Astronomy Centre, University of Sussex, Brighton, UK}

\begin{abstract}
Modelling of the point spread function of the UKIRT IRCAM3 array was conducted in order to test for any extended emission around the X-ray binary Cyg X-3.  We found that the point spread function cannot be represented by a simple Gaussian, but modelling of the stars required additional functions, namely Lorentzian and exponential components.  
After modelling for the PSF, we found that Cyg X-3 could be represented by two stellar-type profiles, 0.56$^{\prime\prime}$ apart.
\end{abstract}
\end{frontmatter}

\section{Introduction}

Cyg X-3 has been studied for 30 years after its discovery in 1966.
During that time, data from the radio, infrared and X-ray wavelengths
has shown it to be a highly unusual source.  This paper concentrates
on the deepest infrared image of Cyg X-3 (Fender \& Bell Burnell,
1996) and subsequent modelling of the profile of Cyg X-3.

The images were taken by Fender using the IRCAM3 array at the UKIRT
telescope in July 1994.  IRCAM3 is a 256 $\times$ 256 InSb imaging
array, operating in the 1 -- 5 $\mu$m wavelength range.  The data here
are composite images from the four bands: H, K, L and J.

\section{H,K,L, J-band images}

\begin{figure}[hbtp]
\begin{picture}(300,460)
\put(-20,450){\includegraphics{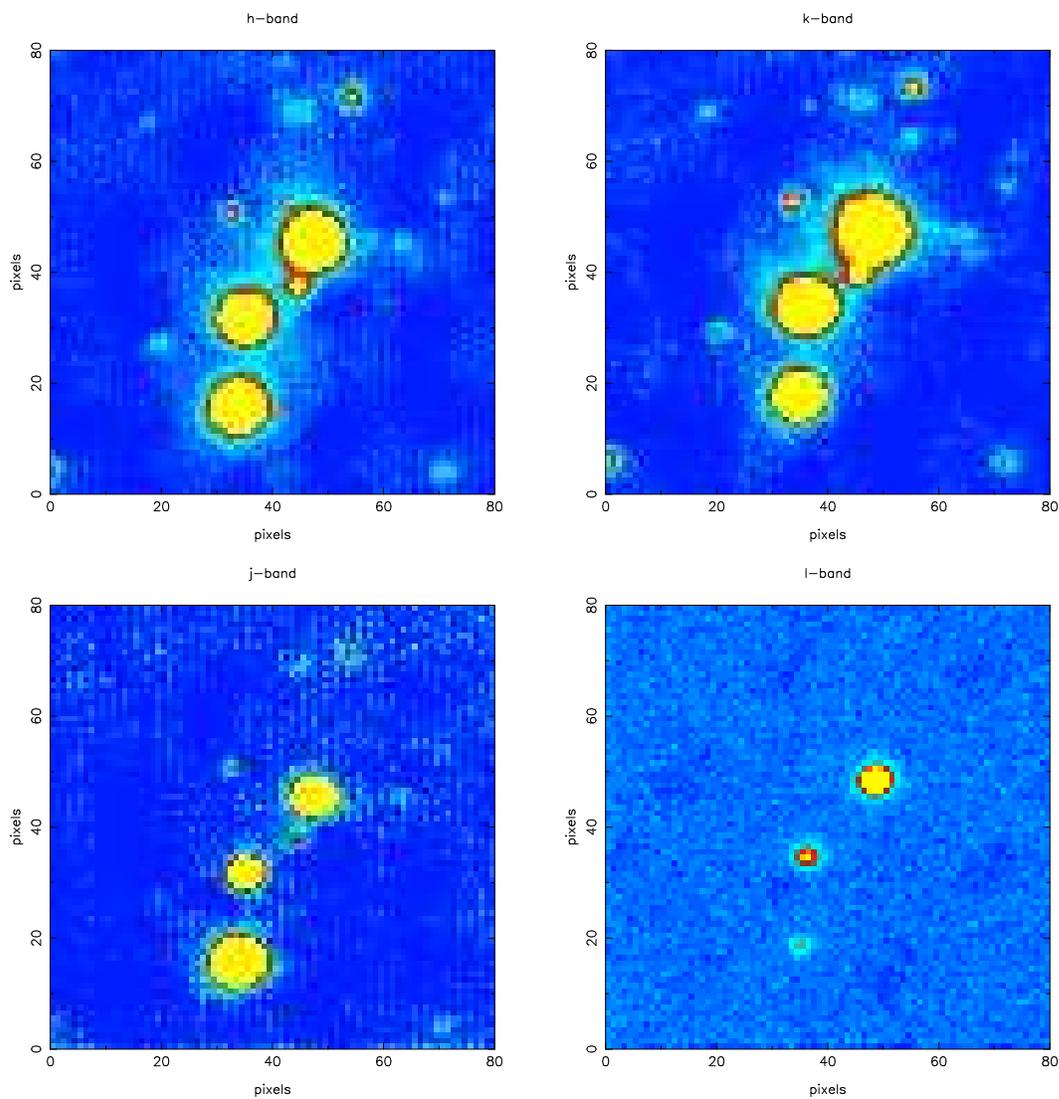}}
\end{picture}
\caption{H,K,L and J-band images of the Cyg X-3 field using the UKIRT IRCAM3 instrument.}
\label{all}
\end{figure}

Figure \ref{all} shows the four bands, H, K, L and J in the
wavelengths 2.2, 1.65, 1.25 and 3.45 $\mu$m respectively.  The images
represent 355 individual frames with integration times between 2 and
25 seconds, binned up for a total integration time of 800 seconds.
Stars in the K-band have fluxes ranging from 12 magnitude (Cyg X-3) to
18th mag for some of the faintest stars.  The positions of the stars
used in this paper were from north to south: Cyg X-3, Star Z and Star
D.  These are the three brightest stars in the images.

\subsection{k-band image}

\begin{figure}[hbtp]
\begin{picture}(300,450)
\put(-30,450){\includegraphics{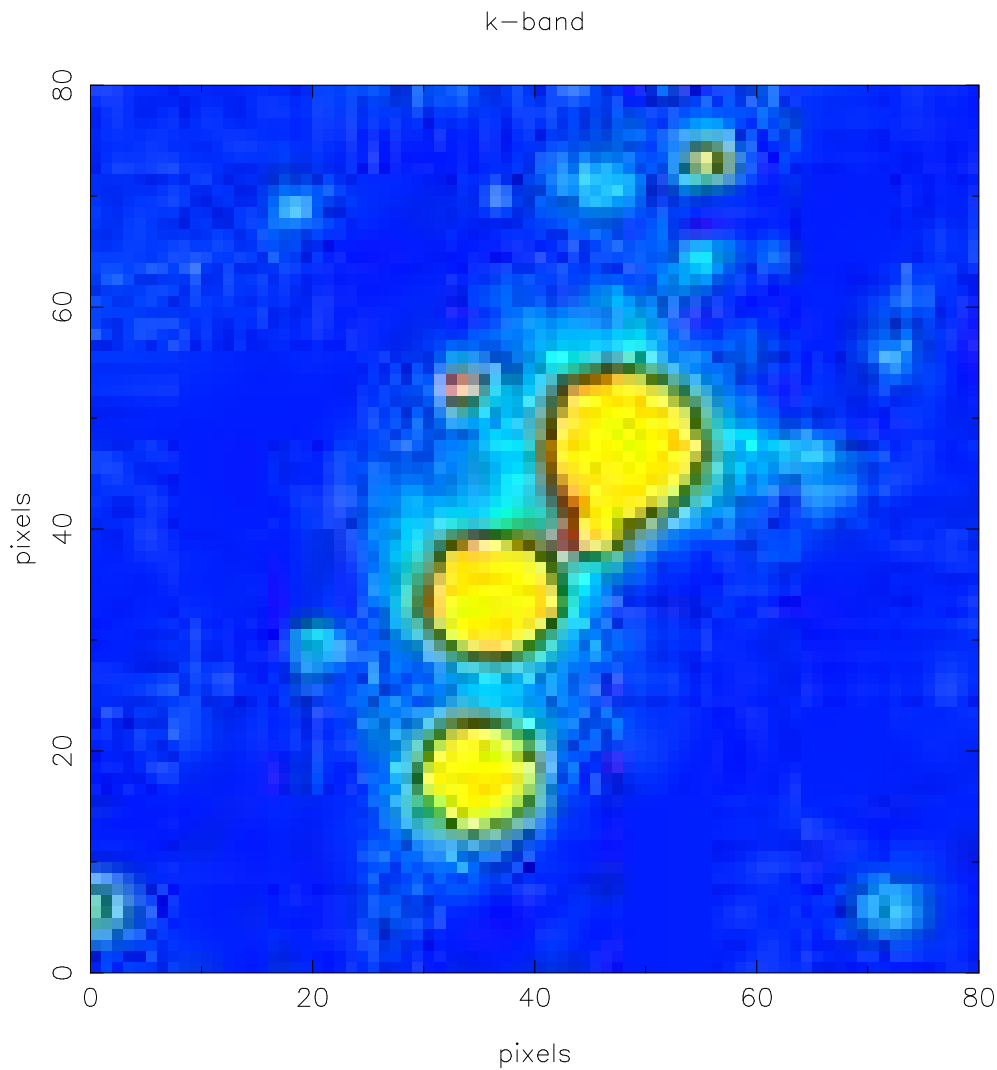}}
\end{picture}
\caption{K-band image}
\label{FendersFinder}
\end{figure}

The k-band image, figure \ref{FendersFinder} or the finder chart
provided by Fender (Fender \& Bell Burnell 1996) was used for all the
modelling done for the IRCAM PSF and the IR emission we are
presenting.  The points of interest are a slight extension to the west
of Cyg X-e and that the image is not quite circular.

\begin{figure}[hbtp]
\begin{picture}(300,260)
\put(-30,290){\includegraphics{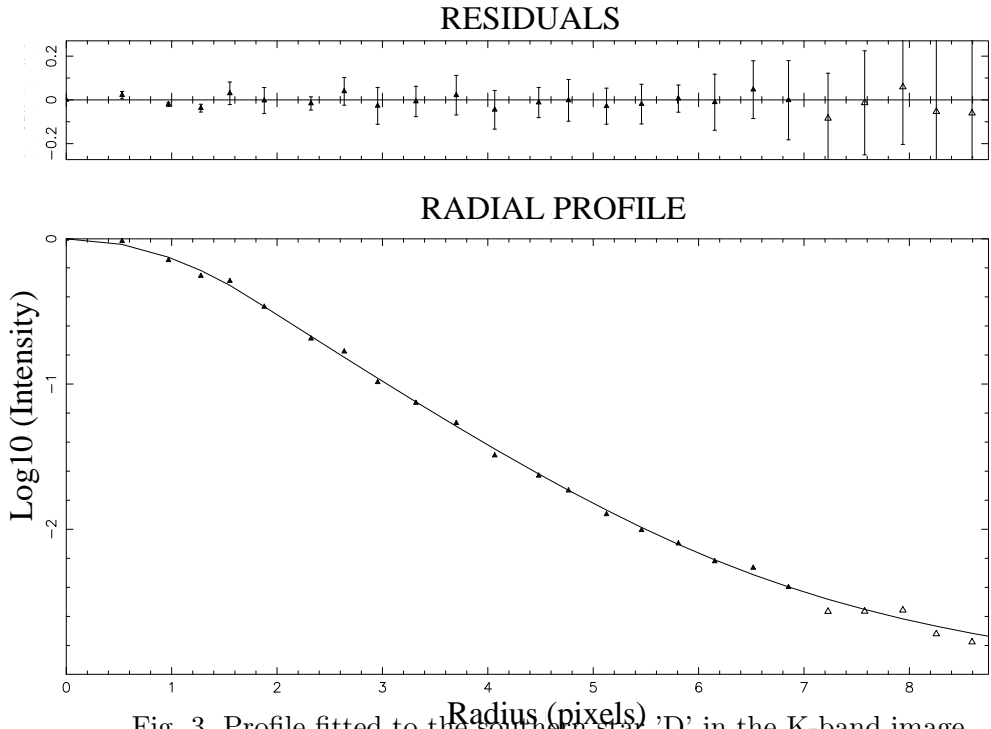}}
\end{picture}
\caption{Profile fitted to the southern star 'D' in the K-band image.}
\label{profile}
\end{figure}

Modelling was then done using the Starlink {\sc pisa} programme.  A
typical stellar profile from this image was obtained using the
southern star, D.  It is important to use a star which is well
separated from the rest of the field as we are important in the faint
extended emission.  This template star has to also be bright enough so
its image covered enough pixels.  Star D was a good compromise between
brightness and separation.  With that as a template {\sc pisa} fitted
a profile as in figure \ref{profile}.  Parameters to the profile are
shown in table \ref{params}, the meaning of which are explained below.

\begin{table}
\begin{tabular}{lc}\hline
Parameter & Value \\ \hline

$\tau$      & 0.32 \\
$\sigma$    & 1.89 \\
$Q$         & 0.066 \\
$R_{\rm c}$ & 2.02 \\ \hline
\end{tabular}
\caption{Parameters fitting the profile shown in figure \ref{profile}.}
\label{params}
\end{table}

These parameters correspond to a set of equations shown below.  You
see that the profile fits very well to the data.  RA wobble in the
telescope is shown in the profile plot by every 3rd point being
slightly brighter that the mean profile.

\section{Profile equations}

The equation governing the profile and also the PSF of the IRCAM3 array can be expressed as:

\begin{eqnarray*}
I_{\rm a} &=& \frac{1}{\pi\sigma^{2}}\left(1+\frac{\tau}{2{\rm ln}\left(1/\tau\right)}\right)^{-1}\left[\frac{Q}{1+r^{2}/\left(\sigma^{2}{\rm ln} 2\right)} + \left(1-Q\right)\exp\left\{\frac{-r^{2}}{\sigma^{2}}\right\}\right]\\ \nonumber
&&{\rm For}\;\;r\leq R_{\rm c}\\
I_{\rm b} &=& \frac{1}{\pi\sigma^{2}}\left(1+\frac{\tau}{2{\rm ln}\left(1/\tau\right)}\right)^{-1}\left[\frac{Q}{1+r^{2}/\left(\sigma^{2}{\rm ln} 2\right)} + \frac{1}{\tau}\left(1-Q\right)\exp\left\{\left(\frac{-2r}{\sigma}\right)\left\{{\rm ln}\left(1/\tau\right)\right\}^{\frac{1}{2}}\right\}\right]\\
&&{\rm For}\;\;r\geq R_{\rm c}\\
\end{eqnarray*}
where
\begin{itemize}
\item[$\tau$]{- Percentage of peak for crossover where the function changes from Gaussian to exponential.}
\item[$\sigma$]{- Gaussian sigma.}
\item[$Q$]{- Percentage of Lorentzian component.}
\item[$R_{\rm c}$]{$= \sigma\left({\rm ln}\left(1/\tau\right)\right)^{\frac{1}{2}}$ - Changeover radius from Gaussian to exponential.}
\end{itemize}

The PSF can be modelled by three parts:

\begin{itemize}
\item[$\rhd$]{Central Gaussian + Lorentzian modification}
\item[$\rhd$]{Exponential tail-off + Lorentzian modification}
\end{itemize}
inside a given radius, $R_{\rm c}$, (2 pixels) the profile is modelled
by a Gaussian with the Gaussian being modified by Lorentzian wings.
Outside a radius of 2 pixels, the profile has an exponential tail-off
with the same Lorentzian wings.

\begin{figure}[hbtp]
\begin{picture}(300,300)
\put(-100,340){\includegraphics{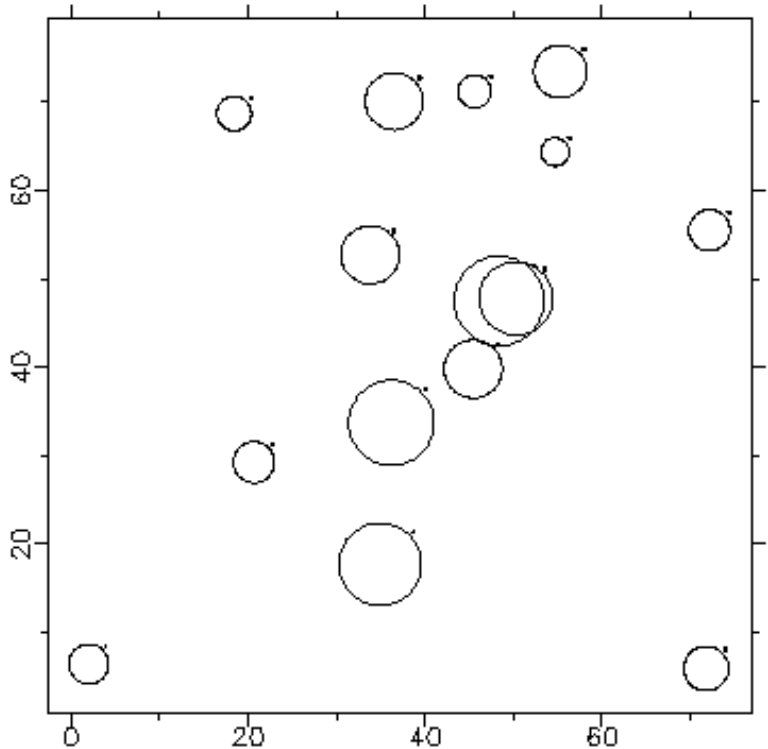}}
\end{picture}
\caption{{\sc pisa} fit for all 'stellar' images to the k-band image using the PSF described above.  Units are pixels, 1 pixel=0.286$^{\prime\prime}$}
\label{overlay}
\end{figure}

Using the profile, {\sc pisa} then searched the k-band image for all
objects that had profiles of the template type and subtracted them
from the image.  After subtracting, {\sc pisa} then re-fitted the
image for any underlying stellar images and figure \ref{overlay} shows
the results of this.

Here {\sc pisa} found emission out to a lower level (represented by
the size of the circles) for all of the stars identified on the finder
chart, except the star to the far west.  We think that is because of
the small dynamic-range for that star and even thought the peak is
similar to other stars found, the background is higher and so it was
not found.

The most interesting thing about this fit is that after re-iterating
the image, {\sc pisa} could fit an additional component to the Cyg X3
image, either in foreground or background.  We are not sure whether
this is a star co-incidental with Cyg X3 or whether it is an artifact
of {\sc pisa}.

\section{Results and conclusions}

This paper attempts to model the UKIRT IRCAM3 point spread function
based on the finder chart and H,L,J-band images.  We find that a
simple Gaussian function is not sufficient in modelling the PSF, and
that one needs to include other functions.  We find that a composite
function of Gaussian with exponential tail-off together with a
Lorentzian modification is sufficient in modelling the PSF.

When the PSF is applied to the K-band finder chart, one can fit two
separate images to the image of Cyg X-3.  These two images are
separated by 0.56$^{\prime\prime}$ and have a flux ratio of 11:1
(left---right).

\section{References}

Fender R.P. \& Bell Burnell S.J., 1996, A\&A, 308, 497\\

\end{document}